\documentclass{article}
\usepackage[preprint]{neurips_2026}

\usepackage{tikz}
\usetikzlibrary{positioning, arrows.meta, calc}
\usepackage{helvet}
\usepackage{xcolor}
\usepackage{tikz}
\usepackage{xcolor}
\usetikzlibrary{arrows.meta, positioning, calc}
\usepackage{graphicx}
\usepackage[utf8]{inputenc}
\usepackage[T1]{fontenc}
\usepackage{hyperref}
\usepackage{url}
\usepackage{booktabs}
\usepackage{amsfonts}
\usepackage{nicefrac}
\usepackage{booktabs}   
\usepackage{longtable}  
\usepackage{amsmath}
\usepackage{microtype}
\usepackage{xcolor}
\usepackage{tikz}
\usetikzlibrary{positioning, arrows.meta, shapes.geometric}

\usepackage[most]{tcolorbox}

\newtcolorbox{positionbox}{
  colback=gray!5,
  colframe=black!55,
  boxrule=0pt,
  leftrule=2.5pt,
  arc=0pt,
  outer arc=0pt,
  left=10pt, right=10pt, top=7pt, bottom=7pt,
  before skip=10pt, after skip=10pt,
}

\newcommand{\corpusN}{438}
\newcommand{\corpusWindow}{2017--2025}
\newcommand{\corpusDetectionN}{389}   

\newcommand{\ReffortPublicN}{276}
\newcommand{\ReffortAudioN}{111}
\newcommand{\ReffortOtherN}{49}
\newcommand{\ReffortNCIIN}{1}
\newcommand{\ReffortMessagingN}{1}
\newcommand{\ReffortRealtimeN}{0}

\newcommand{\ReffortPublic}{71.0}
\newcommand{\ReffortAudio}{28.5}

\newcommand{\benchN}{13}
\newcommand{\benchPublic}{10}
\newcommand{\benchAudio}{3}
\newcommand{\benchNCII}{0}
\newcommand{\benchRealtime}{0}
\newcommand{\benchMessaging}{0}

\newcommand{\harmWindow}{2022--2026}
\newcommand{\harmArup}{\$25\,million}

\title{The Deepfakes We Missed:\\ We Built Detectors for a Threat That Didn't Arrive}

\author{%
  Shaina Raza\thanks{Vector Institute for Artificial Intelligence, Toronto, Canada.} \\
  Vector Institute for Artificial Intelligence \\
  Toronto, ON, Canada \\
  \texttt{shaina.raza@torontomu.ca} \\
}

\begin{document}

\maketitle

\begin{abstract}
Nearly a decade of Machine Learning (ML) research on deepfake detection has been organized around a threat model inherited from 2017--2019, revolving around face-swap and talking-head manipulation of public figures, motivated by concerns about large-scale misinformation and video-evidence fraud. This position paper argues that the threat the field prepared for did not arrive, and the threats that did arrive are substantially different. An accounting of deepfake incidents in 2022--2026 shows that the dominant observed harms are peer-generated Non-Consensual Intimate Imagery (NCII), voice-clone scam calls targeting families and finance workers, and emotional-manipulation fraud. The predicted large-scale public-figure deepfake catastrophe did not materialize during the 2024 global information environment despite extensive preparation. Meanwhile, research effort, benchmarks, and detection methods remain concentrated on the inherited threat model. The central claim of this paper is that this misalignment is now the dominant bottleneck on real-world deepfake defense, not model capability. We argue the ML research community should substantially rebalance its research agenda toward the harm categories that are actually growing. We support this position with empirical accounting of research effort and harm distribution, identify the structural reasons the misalignment persists, and outline three concrete technical research agendas for the under-defended harm categories.
\end{abstract}
 
\section{Introduction}
\label{sec:introduction}
 
We built detectors for a threat that did not arrive. Since 2017, deepfake detection research has been organized around one dominant image of the threat, where a face-swapped or talking-head video of a public figure, deployed at scale to shift an election or break public trust in video evidence.
 Benchmarks \cite{rossler2019faceforensics,li2020celeb,dolhansky2020deepfake}, evaluation protocols, and architectures all inherited this image from the foundational literature of the field \cite{tolosana2020deepfakes,mirsky2021creation}. Nearly a decade later, the threat has not materialized in the form predicted, while different deepfake threats have caused substantial harm without equivalent research attention. We take the \textbf{position} that this misalignment -- not model capability -- is now the dominant constraint on deployable defense, and that the field should rebalance toward the harm distribution that actually materialized.

 \begin{positionbox}
 \vspace{-0.5em}

\textbf{Position.} \textbf{The dominant bottleneck on real-world deepfake defense is no longer model capability but a misalignment between the threat model the field has organized around and the harms that have actually materialized at scale. The ML research community should substantially rebalance its detection research toward three under-defended harm categories: real-time voice-clone detection in telecommunications, on-device privacy-preserving NCII detection with victim-centered workflows, and messaging-layer defenses for peer-distributed synthetic content.}
\vspace{-0.5em}
\end{positionbox}

\begin{figure*}[t]
  \centering
  \includegraphics[width=0.78\textwidth]{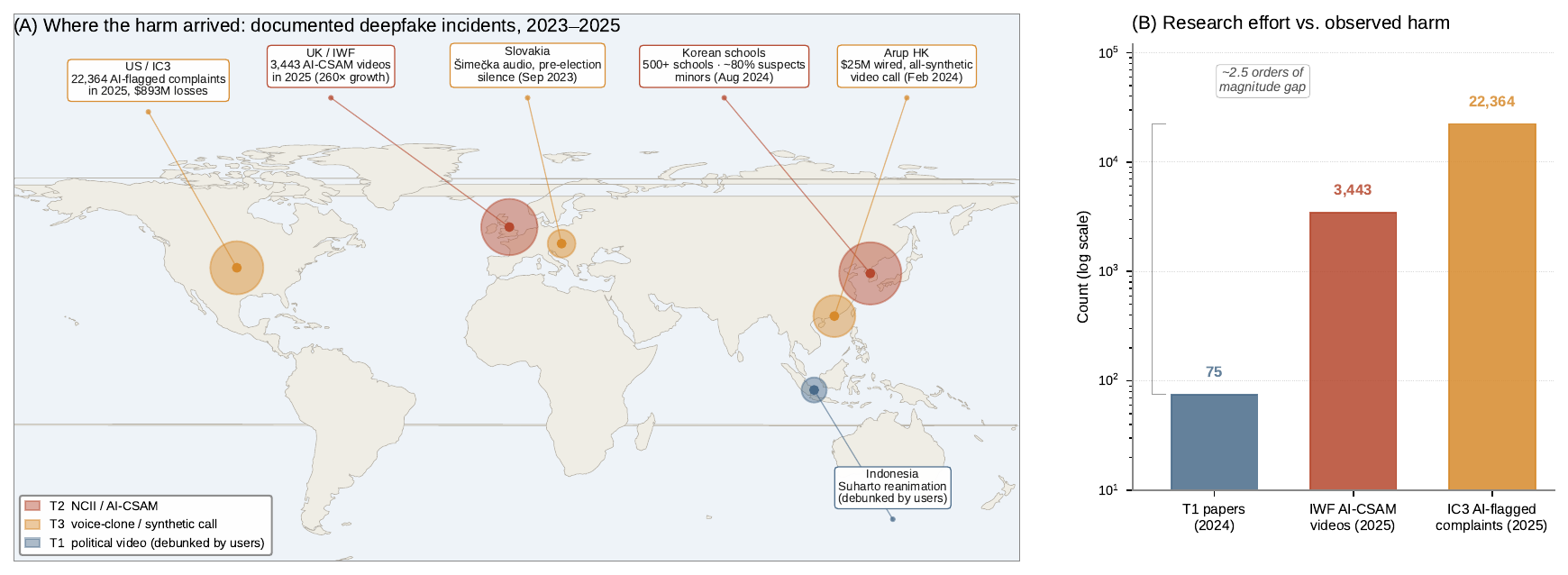}
\caption{\textbf{Research effort by threat category,} \corpusN{}-paper corpus, \corpusWindow{}. T1 (public-figure) dominates every year; under-defended categories are thin slivers (1 T2 in 2025, 1 T5 in 2023, 0 T4). ``Other'' covers surveys, provenance, and unscoped detection methods; \% reported over the \corpusDetectionN{}-paper detection-method subset.}
  \label{fig:harm-gradient}
  \vspace{-1em}
\end{figure*}

\textbf{The predicted political-deepfake catastrophe did not arrive.} Researchers, journalists, and policymakers spent 2022-2024 preparing for a deepfake-driven election disaster. The World Economic Forum 2024 Global Risks Report ranked AI-driven misinformation as the leading global risk over a two-year horizon~\cite{wef_global_risks_2024}, and numerous academic and policy analyses warned of imminent electoral harm~\cite{chesney2019deep}. The 2024 global election cycle, which included national elections in the United States, India, Indonesia, the United Kingdom, and the European Union, produced no documented case in which a deepfake video decisively altered an electoral outcome. Deepfake incidents did occur during these elections, but they were, with few exceptions, identified and debunked by journalists, fact-checkers, and ordinary users rather than by ML detection systems~\cite{simon2024misinformation,panditharatne2024ai}. The threat arrived in smaller, more diffuse forms, and was handled by processes other than those the field was building. Meanwhile, harm at scale arrived elsewhere (for example, peer-generated NCII, voice-clone fraud, and real-time synthetic-identity scams), and the research distribution did not follow, as shown in Figure \ref{fig:harm-gradient} and Section~\ref{sec:misalignment}.
 
\textbf{Scope.}
This paper is not a literature review, a detection method, or a claim that deepfakes are solved or that political deepfakes will never arrive. It is an empirical realignment argument that we compare the distributions of research effort and observed harm across threat categories , diagnose why the misalignment persists, and propose three concrete technical agendas to correct it. A decade of evidence is enough to update research priors.

\textbf{Contributions.} Our contributions are: (1) a 438-paper classification of detection
research \corpusWindow{} across a five-category threat taxonomy (T1--T5, plus Other), showing \ReffortPublic{}\% concentration on T1 public-figure video; (2) a harm distribution synthesized from
IC3~\cite{ic3_reports_fbi}, IWF~\cite{iwf_harm_without_limits_2026}, AIID~\cite{aiid_2026}, StopNCII.org, victim
surveys~\cite{henry2018technology}, and named incidents, concentrated on NCII, voice-clone scams, and peer-distributed manipulation; (3) a diagnosis of why the misalignment persists (benchmark inheritance, dataset-ethics asymmetry, and salience-driven attention); (4) three concrete research agendas for the under-defended
categories, defended on harm-proportionality and ML-tractability.
 
\section{Threat Model Archaeology}
\label{sec:archaeology}
 
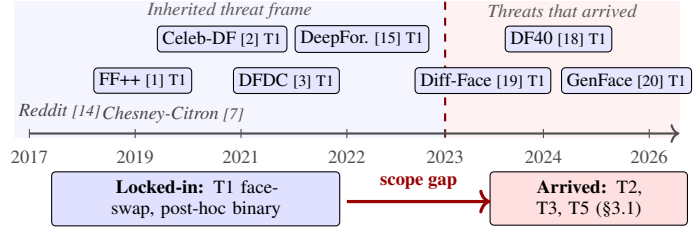
\begin{figure}[t]
\centering
\begin{tikzpicture}[
  scale=1, transform shape,
  font=\footnotesize,
  every node/.style={font=\footnotesize},
  bench/.style={draw, rounded corners=1.5pt, fill=blue!10,
                inner sep=2pt, align=center, font=\scriptsize},
  origin/.style={font=\scriptsize\itshape, text=gray!55!black, align=center},
  assum/.style={draw, rounded corners=1.5pt, fill=blue!12,
                inner sep=3pt, align=center, font=\scriptsize},
  arrived/.style={draw, rounded corners=1.5pt, fill=red!12,
                  inner sep=3pt, align=center, font=\scriptsize},
  era/.style={font=\scriptsize\itshape, text=gray!70!black}
]
\fill[blue!4] (-0.2, 0.3) rectangle (5.5, 2.10);
\fill[red!5]  (5.5,  0.3) rectangle (8.6, 2.10);
\draw[dashed, red!50!black, thick] (5.5, 0.3) -- (5.5, 2.10);
\node[era] at (2.65, 1.97) {Inherited threat frame};
\node[era] at (7.05, 1.97) {Threats that arrived};
\node[bench] (ff)   at (1.50, 1.05) {FF++ {\tiny\cite{rossler2019faceforensics}} \tiny T1};
\node[bench] (cdf)  at (2.55, 1.60) {Celeb-DF {\tiny\cite{li2020celeb}} \tiny T1};
\node[bench] (dfdc) at (3.40, 1.05) {DFDC {\tiny\cite{dolhansky2020deepfake}} \tiny T1};
\node[bench] (deep) at (4.40, 1.60) {DeepFor. {\tiny\cite{jiang2020deeperforensics}} \tiny T1};
\node[bench] (df)   at (6.00, 1.05) {Diff-Face {\tiny\cite{chen2023diffusionface}} \tiny T1};
\node[bench] (df40) at (7.00, 1.60) {DF40 {\tiny\cite{yan2024df40}} \tiny T1};
\node[bench] (gf)   at (7.90, 1.05) {GenFace {\tiny\cite{chen2024genface}} \tiny T1};
\node[origin] at (0.40, 0.65) {Reddit {\tiny\cite{cole2018reddit}}};
\node[origin] at (1.90, 0.60) {Chesney-Citron {\tiny\cite{chesney2019deep}}};
\draw[->, thick, gray!60!black] (0, 0.35) -- (8.6, 0.35);
\foreach \x/\y in {0/2017, 1.4/2019, 2.8/2021, 4.2/2022, 5.5/2023, 6.8/2024, 8.2/2026}
  \draw[gray!60!black] (\x, 0.41) -- (\x, 0.29)
       node[below=1pt, font=\scriptsize]{\y};
\node[assum, text width=3.6cm, anchor=north] at (2.20, -0.15)
  {\textbf{Locked-in:} T1 face-swap, post-hoc binary};
\node[arrived, text width=2.4cm, anchor=north] at (7.40, -0.15)
  {\textbf{Arrived:} T2, T3, T5 (\S\ref{sec:empirical:effort})};
\draw[->, red!60!black, very thick] (4.20, -0.55) -- (6.10, -0.55);
\node[font=\scriptsize\bfseries, text=red!60!black] at (5.15, -0.30) {scope gap};
\end{tikzpicture}
\caption{Threat-model inheritance, 2017--2026. All major benchmarks inherit the T1 face-swap frame seeded by the 2017 Reddit incident and the Chesney-Citron (2018) policy formulation, while the harms that arrived at scale (T2 NCII, T3 voice-clone, T5 peer-distributed; \S\ref{sec:empirical}) sit outside it.}
\label{fig:threat-archaeology}
\vspace{-1em}
\end{figure}

The claim that the field prepared for the wrong threat first requires identifying what threat it did prepare for and how that became the default. This section traces that inheritance (as shown in Fig.~\ref{fig:threat-archaeology}).
 
\paragraph{The origin: 2017--2019}
 
The modern deepfake detection literature originates in late 2017, when a Reddit user under the handle ``deepfakes'' released face-swap videos produced with a consumer autoencoder pipeline, primarily targeting celebrities in non-consensual pornographic content~\cite{cole2018reddit}. This moment fixed both the terminology and the technical paradigm, face-swap video of recognizable individuals, that shaped early research. Two strands then defined the field. On the technical side, FaceForensics++~\cite{rossler2019faceforensics} introduced the dominant benchmark template: curated datasets of manipulated videos with binary detection tasks, later extended by Celeb-DF~\cite{li2020celeb} and the DeepFake Detection Challenge~\cite{dolhansky2020deepfake}. On the policy side, Chesney and Citron 2019 \emph{California Law Review} article~\cite{chesney2019deep} framed deepfakes as threats to privacy, democracy, and national security, illustrated through public-figure scenarios. The two strands were mutually reinforcing: benchmarks defined what counted as a deepfake, and the policy frame defined why it mattered. By 2020, this alignment had hardened into the field default research.
Two mechanisms carried the 2017--2019 threat model forward.
 
\textbf{Benchmark inheritance.}
Subsequent benchmarks (Fig.~\ref{fig:threat-archaeology}) extended generation methods and difficulty but preserved the face-centric, public-figure framing of their predecessors. The core formulation, namely manipulated videos of public figures evaluated via binary classification, went unchallenged across releases spanning GAN-era and early diffusion-era generators~\cite{jiang2020deeperforensics, zhou2021face, le2021openforensics, yan2024df40, chen2023diffusionface, chen2024genface}. As a result, two categories of content remained systematically underrepresented: \emph{peer-generated} content (synthetic media produced by and circulated among non-public individuals, such as classmates, coworkers, or acquaintances) and \emph{private-channel} content (synthetic media distributed through encrypted messaging or closed groups rather than open platforms).

\textbf{Evaluation-protocol lock-in.}
Binary classification at the frame or video level, scored by accuracy measures ~\cite{yan2023deepfakebench, khan2024deepfake}, encodes post-hoc, latency-insensitive, single-decision evaluation as the default.
This inheritance (face-centric data, public-figure framing, and binary post-hoc evaluation) fixed a threat model whose categories (\S\ref{sec:misalignment}) do not match the harms that have since materialized at scale. The remainder of this paper quantifies that gap, diagnoses why it persists, and proposes three research agendas for the categories the inherited frame leaves under-defended.
 
\section{Empirical Misalignment}
\label{sec:misalignment}
\label{sec:empirical}
 
This section evaluates whether the inherited threat model aligns with real-world harms by comparing two distributions: \textbf{research effort}, measured via publication counts, and \textbf{observed harm}, measured through incidents, victims, and financial losses. 
 
\subsection{Research effort by threat category}
\label{sec:empirical:effort}
 
\textbf{Corpus.} We classified \corpusN{} deepfake detection papers
published \corpusWindow{} at major ML, vision, signal-processing, and multimedia venues (e.g., CVPR, AAAI, Interspeech, IEEE TIFS), together with high-citation arXiv preprints. One scope choice is worth flagging: our keyword vocabulary yields sparse coverage at security and HCI venues (USENIX Security, CCS, S\&P, CHI, FAccT), where deepfake work is more often framed as misuse, policy, or platform research than as detection method work. These venues pass our major-venue filter but generate few keyword matches. Full collection pipeline, filters, and a discussion of this limitation are in Appendix~\ref{app:corpus}.
 
Each paper was assigned a primary threat category based on its stated threat model, evaluation datasets, and target use case. The categories are: (i) public-figure face-swap and talking-head video \textbf{(T1)}; (ii) peer-generated NCII \textbf{(T2)}; (iii) audio / voice-clone \textbf{(T3)}; (iv) real-time / live-stream \textbf{(T4)}; (v) messaging-layer / peer-distributed \textbf{(T5)}; and (vi) Other.

Two grouping choices are worth flagging. We treat face-swaps of public figures and talking-head synthesis as a single category (T1) because their benchmarks, generators, and evaluation pipelines overlap heavily, while noting that talking-head methods can  target private faces. We report T5 separately from T2 and T3 even though it is best understood as a distribution channel for them, because the channel constraints (encryption, peer-distribution, on-device processing) drive different defensive requirements.
A paper is assigned to T2, T4, or T5 when the threat object is the dominant focus of its title and abstract under our weighted keyword rules, rather than a peripheral reference; ties at category boundaries break against T1 rather than toward it. Full classification rules, keyword filters, confidence stratification, and the adversarial-reassignment  robustness check are in App.~\ref{app:classification}.
 
\begin{figure}[t]
  \centering
  \includegraphics[width=0.7\linewidth]{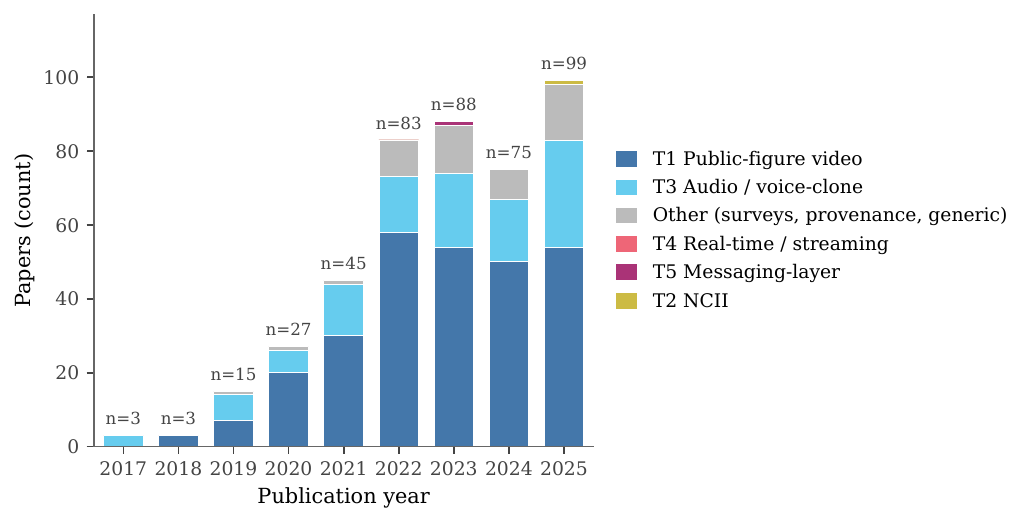}
\caption{Research effort by threat category, \corpusN{}-paper corpus, \corpusWindow{}. T1 (public-figure) dominates every year; the under-defended categories are thin slivers where present at all (1 T2 paper in 2025, 1 T5 in 2023, 0 T4 across the window). ``Other'' covers surveys, provenance, and detection methods not scoped to a named threat; \% in the text are over the \corpusDetectionN{}-paper detection-method subset (corpus minus Other).}
  \label{fig:effort}
  \vspace{-1em}
\end{figure}

\textbf{Distribution.} The distribution is concentrated. Of the
\corpusN{}-paper corpus, \ReffortOtherN{} papers (11.2\%) fall outside the named threat categories. This residual is heterogeneous and worth unpacking, as it includes surveys, generic detection methods not scoped to a named threat object, and a substantial provenance and watermarking literature. 
The remaining \corpusDetectionN{} papers address a specific threat category; we report shares over this detection-method subset to keep the comparison with benchmarks (which all target specific threat objects) meaningful. Within this subset, \ReffortPublic{}\% (\ReffortPublicN{} papers) target T1. T3 (audio/voice-clone) accounts for a further \ReffortAudio{}\% (\ReffortAudioN{} papers), but the real-time/telephony-degraded subset relevant to consumer attacks (\S\ref{sec:observed-harm}) is essentially absent. The 28.5\% T3 share is almost entirely offline ASVspoof-style  \cite{wu2017asvspoof} anti-spoofing on pre-recorded utterances, even where telephony codecs are included as a robustness axis. The 0\% T4 figure refers to the distinct consumer-scam setting: sub-second alerting on streaming audio with speaker-conditional verification, which existing T3 infrastructure does not cover. The three under-defended categories together return fewer than five papers under our primary-evaluation rule: \ReffortNCIIN{} for T2, \ReffortMessagingN{} for T5, \ReffortRealtimeN{} for T4. The shape is what matters: T1 dominant every year, T3 substantial but evaluation-mismatched, T2/T4/T5 thin to absent (Figure~\ref{fig:effort}). T1 dominance has  strengthened as foundation-model generators have expanded the attack surface for the same threat object.
 
\textbf{Benchmark coverage tracks paper distribution.} We classified the \benchN{} widely cited deepfake benchmarks under the same schema. \benchPublic{} of \benchN{} target public-figure face or
talking-head video; \benchAudio{} target audio or audio-visual settings (FakeAVCeleb~\cite{khalid2021fakeavceleb}, LAV-DF~\cite{cai2023lavdf}, AV-Deepfake1M~\cite{cai2024avdeepfake1m} are the principal examples); \benchNCII{} target peer-generated NCII; and \benchRealtime{} target
real-time or streaming conditions. Full table in Appendix~\ref{app:benchmarks}. The benchmark distribution is not just similar to the paper distribution; it shapes it. Methods are trained and evaluated on what benchmarks exist, and the absence of T2, T4, and T5 benchmarks is a structural constraint on what work the field can recognize as a contribution.
 
\subsection{Observed harm by threat category}
\label{sec:observed-harm}
\label{sec:empirical:observed}
 
We synthesize harm data from five public sources covering \harmWindow{}: the FBI Internet Crime Complaint Center annual reports~\cite{ic3_reports_fbi}, the Internet Watch Foundation's AI-generated CSAM monitoring~\cite{iwf_harm_without_limits_2026}, the AI Incident Database~\cite{aiid_2026}, academic victim-prevalence surveys~\cite{henry2018technology}, and documented high-profile incidents reported in reputable news outlets. No single source is comprehensive; none is free of reporting bias. Taken together, they establish the direction and approximate magnitude of the harm distribution, which is the level of claim this paper requires.(discussion appears in Appendix~\ref{app:harm-methodology}).
 
\textbf{Peer-generated NCII.} The Internet Watch Foundation reports sustained year-over-year growth in AI-generated child sexual abuse imagery, with videos rising from 13 in 2024 to 3{,}443 in 2025 (a 260-fold increase), and 8{,}029 images and videos assessed in total~\cite{iwf_harm_without_limits_2026}. The August 2024 South Korean school deepfake incident involved more than 500 schools and thousands of victims, with over 80\% of identified suspects being minors~\cite{bbc2024koreadf,eastasia2024koreadf}. Distribution occurred mainly via encrypted messaging platforms (Telegram-class group chats), indicating that T5 is better understood as a channel through which T2 (and, in voice-clone settings, T3) harm propagates rather than as an independent threat. The defensive focus shifts from post-hoc media detection to intervention at distribution and on-device endpoints. StopNCII.org and similar platforms process submissions at population scale. This is the largest observed harm category by victim count, yet the research literature (under 1\% of the corpus) remains limited.
 
\textbf{Voice-clone and real-time synthetic-identity fraud.} The FBI Internet Crime Complaint Center has documented rapid growth in synthetic-media-enabled fraud, with reported losses in the billions~\cite{ic3_reports_fbi}. The 2024 Arup case, in which a finance employee authorized a \harmArup{} transfer after a video call in which every other participant was a synthetic avatar, is the most widely reported instance~\cite{cnn_deepfake_cfo_2024}, but consumer-facing voice-clone scams targeting families (grandparent scams, kidnapping-ransom hoaxes) account for a larger share of reported incidents by volume. The attack surface is conversational audio delivered over telephony, often with latency requirements that preclude offline analysis. 

 
\textbf{Public-figure political video.} The 2024 global election cycle (US, India, Indonesia, UK, EU national elections) produced no documented case in which a synthetic political \emph{video} was the post-election forensic verdict on a result change. Video incidents did occur, including the Indonesian Suharto reanimation, and several drew substantial media attention; the cases documented by fact-checkers were overwhelmingly identified by journalists and ordinary users, not by deployed ML detection. We restrict the claim to video deliberately: it is the modality the detection field has overwhelmingly prepared for, and so the modality on which the prediction-versus-outcome comparison is most diagnostic. The most-cited candidate counterexample (the September 2023 Slovakia \v{S}ime\v{c}ka recording, released during pre-election media silence) sits in T3 rather than T1, and the broader pattern is consistent: where political synthetic media has plausibly affected outcomes, it has done so in audio. The category the field built for is not the category in which harm has materialized at scale, and the closest counterexamples sit in another under-defended category.
\begin{figure}[t]
  \centering
  \includegraphics[width=0.85\linewidth]{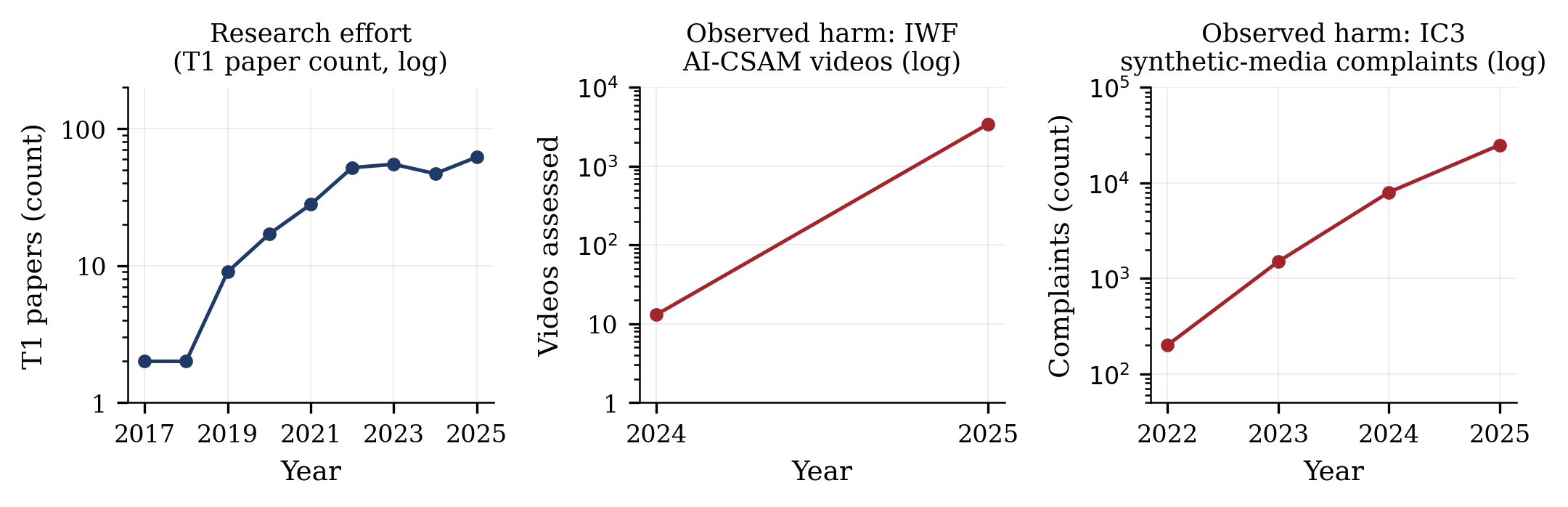}
\caption{Research effort vs. observed harm, 2017--2025, on a common log scale. Left: T1 (public-figure) paper count per year in the \corpusN-paper corpus. Centre: AI-generated CSAM videos assessed by IWF~\cite{iwf_harm_without_limits_2026}. Right: synthetic-media-flagged complaint count from IC3 annual reports~\cite{ic3_reports_fbi} (we use complaints rather than dollar losses to avoid attribution ambiguity from BEC aggregation; see Appendix~\ref{app:harm-methodology}). The claim is the divergence of slopes: T1 paper volume grows in a linear band, while the harm indicators jump multiple orders of magnitude over a comparable window.}
\label{fig:stability}
\vspace{-1em}
\end{figure}
\subsection{The misalignment is stable and directional}
\label{sec:empirical:stable}
 
Two features of the comparison matter for the position this paper takes.
First, the misalignment is not a single-year artifact. Figure~\ref{fig:effort} compares absolute counts on a common log scale: T1 paper volume rises in a linear band over 2017--2025 (a roughly 30-fold increase from the 2017--2018 baseline), while the two harm indicators rise much faster (IWF AI-CSAM video assessments grow 260-fold between 2024 and 2025 alone~\cite{iwf_harm_without_limits_2026}, and IC3 synthetic-media-flagged complaints grow more than an order of magnitude across the available reporting window~\cite{ic3_reports_fbi}).
 The gap between the distributions has been widening, not narrowing.
Second, the misalignment is directional and not reducible to measurement bias. Harm-side undercounting (peer-generated NCII in encrypted channels, voice-clone fraud absorbed into BEC statistics) makes the gap worse, not better: the categories the field has prepared for are not the undercounted ones. Research-side overcounting via cross-category transfer is also unsupported , transfer evidence is essentially absent in the corpus (Appendix~\ref{app:transfer}). Why this distribution has persisted despite being visible for years (Section~\ref{sec:persistence}) determines whether rebalancing is achievable or merely advisable.
 
\section{Why the Misalignment Persists}
\label{sec:persistence}
 
The misalignment in Section~\ref{sec:empirical} has been visible for years but the research has not shifted, sustained by a stable equilibrium of three reinforcing factors.

\textbf{Benchmark inheritance as publication gravity}
\label{sec:persist-benchmarks}
For a working researcher, benchmarks are not primarily a source of data but a source of publication legibility. A method evaluated on FaceForensics++, Celeb-DF, and DFDC enters review with an implicit comparator set, a familiar metric (AUC on frame- or video-level classification), and a clear claim structure. A method evaluated on a custom corpus of voice-clone scam calls, or peer-generated images under restricted access, must justify its evaluation from first principles.
 
The asymmetry is rational for reviewers calibrating on comparability and for authors shipping legible contributions under compressed cycles. The consequence is that inherited benchmarks exert gravitational pull largely independent of whether the threat object they encode is still the most harmful. A graduate student in 2025 finds public-figure deepfake detection legible by default; real-time voice-clone telephony detection is legible only after substantial overhead in dataset construction, protocol justification, and comparator search. Researchers are not unaware of the alternatives; the publication system rewards the direction the benchmarks point, and they point at the 2017--2019 threat model.
 
\textbf{Dataset ethics as an asymmetric tractability gradient}
\label{sec:persist-ethics}
The largest-harm categories are those where the standard research pipeline is foreclosed. NCII corpora cannot be shared under face-manipulation norms; AI-generated CSAM (child sexual abuse material) cannot be collected at all; voice-clone scam recordings sit with telecoms and law enforcement. Tractability runs opposite to harm: face-manipulation data is abundant and corpus-level uncomplicated, with small victim counts; NCII and CSAM data is scarce and legally constrained, with large victim counts.
This does not make NCII or CSAM research impossible. It means the model behind the face-manipulation literature (open corpora, shared benchmarks, dataset reuse) does not port, and no replacement exists. Victim-centered workflows, privacy-preserving evaluation, platform and hotline partnerships, and federated benchmarking are partial answers, all underdeveloped. 
 
\textbf{Media and policy salience as the field input signal}
\label{sec:persist-salience}
Attention is allocated by salience rather than incidence. Political deepfakes generate high per-incident coverage: a single manipulated video of a head of state produces international reporting, policy hearings, and funding calls. Peer-generated NCII and voice-clone fraud generate low per-incident coverage, aggregated into periodic summary reports from IC3, IWF, and victim-support organizations. The per-incident ratio is inverted relative to the aggregate-harm ratio, and the field tracks the per-incident signal. Funding calls, workshop themes, and special issues follow the salient threat, and press coverage of ML research rewards papers that address it because the framing is pre-built.
 
\section{Three Research Agendas for the Under-Defended Harm Categories}
\label{sec:agendas}

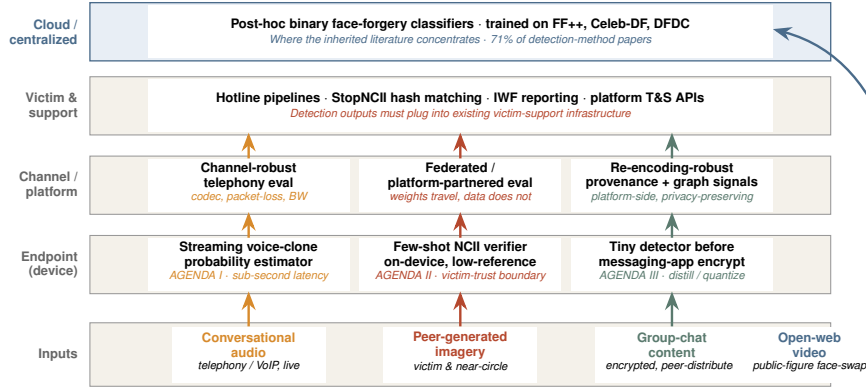
\begin{figure*}[t]
\centering

\definecolor{tone}{HTML}{4A6B8A}
\definecolor{ttwo}{HTML}{B5482E}
\definecolor{tthree}{HTML}{D78A2C}
\definecolor{tfive}{HTML}{5A7A6E}
\definecolor{tierfill}{HTML}{F4F1EA}
\definecolor{inhfill}{HTML}{E8EEF5}

\begin{tikzpicture}[
    scale=0.7, transform shape,
    font=\sffamily\footnotesize,
    tier/.style={
      rectangle, draw=black!40, line width=0.4pt,
      fill=tierfill, minimum width=14cm, minimum height=1.1cm
    },
    inhtier/.style={
      rectangle, draw=tone, line width=0.5pt,
      fill=inhfill, minimum width=14cm, minimum height=1.1cm
    },
    tierlabel/.style={
      font=\sffamily\bfseries\scriptsize, text=black!60,
      align=right, anchor=east
    },
    compbase/.style={
      rectangle, draw=none, fill=white,
      align=center, inner sep=2pt
    },
    heading/.style={
      font=\sffamily\bfseries\scriptsize, align=center
    },
    sub/.style={
      font=\sffamily\itshape\tiny, align=center
    },
    flow/.style={->, line width=0.9pt, >=Stealth},
    sweep/.style={->, line width=0.9pt, >=Stealth, rounded corners=6pt}
]

\node[inhtier] (cloudband)    at (0, 5.4) {};
\node[tier]    (victimband)   at (0, 4.0) {};
\node[tier, minimum height=1.1cm]  (channelband)  at (0, 2.5) {};
\node[tier, minimum height=1.1cm]  (endpointband) at (0, 1.0) {};
\node[tier, minimum height=1.2cm]  (inputband)    at (0,-0.7) {};

\node[tierlabel, text=tone]  at ($(cloudband.west)+(-0.1,0)$)   {Cloud /\\centralized};
\node[tierlabel]             at ($(victimband.west)+(-0.1,0)$)   {Victim \&\\support};
\node[tierlabel]             at ($(channelband.west)+(-0.1,0)$)  {Channel /\\platform};
\node[tierlabel]             at ($(endpointband.west)+(-0.1,0)$) {Endpoint\\(device)};
\node[tierlabel]             at ($(inputband.west)+(-0.1,0)$)    {Inputs};

\node[compbase, minimum width=11.8cm, minimum height=1cm]
  (cloudbox) at (0, 5.4) {};
\node[heading] at ($(cloudbox.center)+(0,0.13)$)
  {Post-hoc binary face-forgery classifiers $\cdot$ trained on FF++, Celeb-DF, DFDC};
\node[sub, text=tone] at ($(cloudbox.center)+(0,-0.16)$)
  {Where the inherited literature concentrates $\cdot$ 71\% of detection-method papers};

\node[compbase, minimum width=11.8cm, minimum height=1cm]
  (victimbox) at (0, 4.0) {};
\node[heading] at ($(victimbox.center)+(0,0.13)$)
  {Hotline pipelines $\cdot$ StopNCII hash matching $\cdot$ IWF reporting $\cdot$ platform T\&S APIs};
\node[sub, text=ttwo] at ($(victimbox.center)+(0,-0.16)$)
  {Detection outputs must plug into existing victim-support infrastructure};

\node[compbase, minimum width=3.8cm, minimum height=1.0cm]
  (chan0) at (-4.0, 2.5) {};
\node[heading] at ($(chan0.center)+(0,0.18)$)
  {Channel-robust\\telephony eval};
\node[sub, text=tthree] at ($(chan0.center)+(0,-0.22)$)
  {codec, packet-loss, BW};

\node[compbase, minimum width=3.8cm, minimum height=1.0cm]
  (chan1) at (0, 2.5) {};
\node[heading] at ($(chan1.center)+(0,0.18)$)
  {Federated /\\platform-partnered eval};
\node[sub, text=ttwo] at ($(chan1.center)+(0,-0.22)$)
  {weights travel, data does not};

\node[compbase, minimum width=3.8cm, minimum height=1.0cm]
  (chan2) at (4.0, 2.5) {};
\node[heading] at ($(chan2.center)+(0,0.18)$)
  {Re-encoding-robust\\provenance + graph signals};
\node[sub, text=tfive] at ($(chan2.center)+(0,-0.22)$)
  {platform-side, privacy-preserving};

\node[compbase, minimum width=3.8cm, minimum height=1.0cm]
  (ep0) at (-4.0, 1.0) {};
\node[heading] at ($(ep0.center)+(0,0.18)$)
  {Streaming voice-clone\\probability estimator};
\node[sub, text=tthree] at ($(ep0.center)+(0,-0.22)$)
  {AGENDA I $\cdot$ sub-second latency};

\node[compbase, minimum width=3.8cm, minimum height=1.0cm]
  (ep1) at (0, 1.0) {};
\node[heading] at ($(ep1.center)+(0,0.18)$)
  {Few-shot NCII verifier\\on-device, low-reference};
\node[sub, text=ttwo] at ($(ep1.center)+(0,-0.22)$)
  {AGENDA II $\cdot$ victim-trust boundary};

\node[compbase, minimum width=3.8cm, minimum height=1.0cm]
  (ep2) at (4.0, 1.0) {};
\node[heading] at ($(ep2.center)+(0,0.18)$)
  {Tiny detector before\\messaging-app encrypt};
\node[sub, text=tfive] at ($(ep2.center)+(0,-0.22)$)
  {AGENDA III $\cdot$ distill / quantize};

\node[compbase, minimum width=2.9cm, minimum height=1.05cm]
  (in0) at (-4.0, -0.7) {};
\node[heading, text=tthree] at ($(in0.center)+(0,0.18)$)
  {Conversational\\audio};
\node[sub] at ($(in0.center)+(0,-0.22)$)
  {telephony / VoIP, live};

\node[compbase, minimum width=2.9cm, minimum height=1.05cm]
  (in1) at (0, -0.7) {};
\node[heading, text=ttwo] at ($(in1.center)+(0,0.18)$)
  {Peer-generated\\imagery};
\node[sub] at ($(in1.center)+(0,-0.22)$)
  {victim \& near-circle};

\node[compbase, minimum width=2.9cm, minimum height=1.05cm]
  (in2) at (4.0, -0.7) {};
\node[heading, text=tfive] at ($(in2.center)+(0,0.18)$)
  {Group-chat\\content};
\node[sub] at ($(in2.center)+(0,-0.22)$)
  {encrypted, peer-distributed};

\node[compbase, minimum width=2.9cm, minimum height=1.05cm]
  (in3) at (6.6, -0.7) {};
\node[heading, text=tone] at ($(in3.center)+(0,0.18)$)
  {Open-web\\video};
\node[sub] at ($(in3.center)+(0,-0.22)$)
  {public-figure face-swap};

\draw[flow, draw=tthree] (in0.north) -- (ep0.south);
\draw[flow, draw=ttwo]   (in1.north) -- (ep1.south);
\draw[flow, draw=tfive]  (in2.north) -- (ep2.south);

\draw[flow, draw=tthree] (ep0.north) -- (chan0.south);
\draw[flow, draw=ttwo]   (ep1.north) -- (chan1.south);
\draw[flow, draw=tfive]  (ep2.north) -- (chan2.south);

\draw[flow, draw=tthree] (chan0.north) -- ($(victimbox.south -| chan0)$);
\draw[flow, draw=ttwo]   (chan1.north) -- ($(victimbox.south -| chan1)$);
\draw[flow, draw=tfive]  (chan2.north) -- ($(victimbox.south -| chan2)$);

\draw[->, line width=0.9pt, draw=tone, >=Stealth]
  ($(in3.north east)+(0.05,0)$)
  to[out=75, in=-15] ($(cloudbox.east)+(0,0.05)$);

\end{tikzpicture}

\caption{Defensive architecture for deepfake harm. Inputs (bottom) reach
the user through one of four tiers; the inherited literature occupies the
cloud tier and addresses only the open-web public-figure video input,
while the three agendas of \S\ref{sec:agendas} defend at the endpoint,
channel, and victim/support tiers, addressing the input contexts (T2,
T3, T5) the inherited tier does not cover.}
\label{fig:architecture}
\vspace{-1em}
\end{figure*}
Sections~\ref{sec:empirical}--\ref{sec:persistence} identified three high-incidence harm categories where current methods do not transfer. We outline an agenda for each and specify the harm, the open technical problems, and the evaluation protocols required.

\paragraph{Agenda I: Real-time voice-clone detection for telecommunications}
\label{sec:agenda-audio}
\emph{The harm.} Voice-clone scam calls are the fastest-growing deepfake
harm category by reported financial loss~\cite{ic3_reports_fbi}. The
attack surface is conversational audio over telephony or VoIP, exploiting family relationships (grandparent scams, kidnapping-ransom hoaxes) or authority structures (CFO impersonation, vendor-payment fraud); the 2024 Arup case~\cite{cnn_deepfake_cfo_2024} illustrates the upper end.
\emph{Open problems.} Three are central. \textbf{(i) Latency-constrained
streaming detection.} A running synthesis-probability estimate must be produced within a conversational alerting budget; window size, update frequency, and false-positive rates are under-addressed today.
\textbf{(ii) Channel-robust detection.} Telephony codecs, bandwidth
restriction, and packet-loss concealment distort the artifacts offline
detectors rely on, so evaluation must use realistic telephony degradation rather than clean-channel audio. \textbf{(iii) Speaker-conditional detection.} Whether audio is the specific person it claims to be, given reference samples, rames a one-shot verification problem layered on synthesis detection , a natural bridge to speaker-verification work that current literature underuses. 

\emph{Evaluation protocols.} Benchmarks need realistic telephony channels
(not clean-studio recordings), decision-latency metrics alongside accuracy, and social-engineering conditions rather than long read-aloud prompts. None exist today; building one is itself a contribution.
\paragraph{Agenda II: On-device and privacy-preserving NCII detection}
\label{sec:agenda-ncii}
\emph{The harm.} Peer-generated NCII, including AI-generated CSAM, is
the largest observed deepfake harm category by victim
count~\cite{iwf_harm_without_limits_2026,bbc2024koreadf}. Victims are overwhelmingly private individuals with scarce reference imagery and severely constrained disclosure contexts. The harm is not whether a
detector identifies a manipulation post hoc; it is whether victims have
workable recourse, whether platforms can intervene at distribution,
and whether perpetrators are deterred.
\emph{Open problems.} \textbf{(i) Low-reference verification.} A victim
has a few consensual reference images and a larger candidate set to
triage, posing a few-shot learning problem sitting between
identity-agnostic detection and full identity enrolment.
\textbf{(ii) On-device and privacy-preserving detection.} The decision
often needs to happen without candidate images leaving the victim's or
platform's trust boundary; federated, on-device, and cryptographic
methods developed elsewhere in ML are under-applied here.
\textbf{(iii) Victim-centered workflow integration.} Detection outputs
must plug into existing victim-support and moderation infrastructure
(StopNCII.org hash matching, platform trust-and-safety APIs, national
hotlines), and detection errors propagate through these pipelines with
asymmetric costs.
\emph{Evaluation protocols.} The ``curate corpus, release publicly,
invite methods'' template does not apply. Compatible alternatives
include \emph{federated evaluation} (weights travel, data does not),
\emph{platform-partnered evaluation} (metrics computed inside a
partner's trust boundary), and \emph{synthetic-victim evaluation}
(controlled-condition images with consenting adult participants, as a
lower-bound benchmark before partner data).

\paragraph{Agenda III: Messaging-layer defenses for peer-distributed synthetic content}
\label{sec:agenda-messaging}
\emph{The harm.} The Korean school
incident~\cite{bbc2024koreadf,eastasia2024koreadf} exemplifies a
category defined not by generation quality or detector sophistication
but by \emph{distribution}: peer-produced imagery circulating through
group chats, often on encrypted channels, at a scale centralized
detection cannot reach.
\emph{Open problems.} \textbf{(i) On-device endpoint detection.} Small
models run on user devices to process media before encryption under
constrained compute and memory, inverting the generous-compute setting
that current literature evaluates; distillation, quantization, and
architectural search for tiny detection models are priority directions.
\textbf{(ii) Provenance robust to re-encoding.} Messaging platforms
re-encode on upload and degrade or destroy watermarks designed for web
distribution, requiring robust provenance signals composed with
detection methods that fall back when signals are absent.
\textbf{(iii) Graph and distribution-pattern signals.} Platform-side
signals (re-upload frequency, cross-group propagation, sender-recipient
graph) can be exploited under privacy-preserving constraints alongside
endpoint detection.
\emph{Evaluation protocols.} Three benchmark types are needed and none
exist today: \emph{endpoint-latency} (decisions under device compute
constraints), \emph{re-encoding-robust} (media passed through realistic
platform encoding pipelines), and \emph{distribution-graph}
(synthetic-content spread simulated over realistic sender-recipient
graphs). 

\noindent A marginal unit of effort on any of these agendas would reduce
observed harm more than the same unit spent on the inherited threat
model; that is the operational form of this paper's position.
\section{Alternative Positions}
\label{sec:alternatives}

A reviewer could grant the empirical evidence of Section~\ref{sec:empirical} and still reject the rebalancing argument. We engage three such positions.
 
\textbf{Alternative View 1: Capability-overhang consolidation}
\label{sec:alt-capability}
 Foundation-model generators have improved faster than detectors, and the gap is widening. The right response is to concentrate scarce detection effort on the hardest generator class, such as high-quality public-figure video from large foundation models. A detector handling state-of-the-art face-swap will handle weaker generators used for NCII and voice-clone scams as a corollary. Consolidation on the hardest case has historically worked in image classification and speech recognition.\\
 \textbf{Response.} 
The transfer claim is empirically weak in the deepfake-detection case. Cross-dataset generalization \textit{within} the public-figure category already shows substantial performance drops; evidence for transfer \textit{across} categories (video-to-audio, public-to-private, offline-to-real-time) is essentially absent in our corpus (Section~\ref{sec:empirical:stable}, Appendix~\ref{app:transfer}). Consolidation works where a single evaluation axis exists; deepfake harm is distributed across axes (modality, subject, channel, latency) that are not reducible to one another.

\textbf{Alternative View 2: Provenance-over-detection}
\label{sec:alt-provenance}
Detection is a losing paradigm in the long run: generators will keep improving, detector-generator arms races favor the generator, and the correct long-term defense is cryptographic provenance (C2PA and successors), generation-time watermarking, and authenticated-capture infrastructure. Rebalancing deepfake research means shifting effort \emph{out of} detection, not redistributing it across categories. \\
\textbf{Response.} Even fully successful, provenance does not cover the harms of Section~\ref{sec:empirical:observed}. It answers ``was this media produced by an authenticated source''; it does not answer ``is this audio stream on my phone call my daughter.'' Voice-clone scam calls are live conversational audio with no sender-side signature to verify. Peer-generated NCII is produced by consumer generators with no incentive to embed signals, distributed over channels that strip them if present, and consumed in contexts where the victim's workflow is not a signature-verification workflow. Provenance covers distribution of authenticated media; it does not cover the channels where the documented harm sits. Provenance and detection are complementary rather than substitutable, and Section~\ref{sec:agenda-messaging} explicitly includes provenance components where they apply.
 
\textbf{Alternative View 3: The harm-is-not-a-deepfake-problem position}
\label{sec:alt-reframing}
 The harms of Section~\ref{sec:empirical:observed} are not primarily deepfake problems: voice-clone scams succeed via social engineering, peer-generated NCII via platform distribution dynamics and coercive social conditions. The effective interventions are platform governance, user education, and law-enforcement, none of which is ML research.\\
\textbf{Response.} Synthetic-media generation is a force multiplier that has measurably changed the scale and victim demographics of each category: voice-clone scams at current volume did not exist before consumer voice tools, and peer-generated NCII at the Korean-incident scale~\cite{bbc2024koreadf,eastasia2024koreadf} was not possible before consumer image generation reached school-aged users. The IWF AI-CSAM growth~\cite{iwf_harm_without_limits_2026} is a direct measurement of synthetic-media-enabled harm not reducible to background distribution . 
 
 
\section{Objections, Limitations, and a Call}
\label{sec:conclusion}
 
A position paper of this shape invites specific objections. We address the two we find most substantive:
 
\subsection{Objections}
\label{sec:objections}
 
\textbf{Objection 1: The political deepfake catastrophe has not arrived yet; precaution is warranted.} This is the strongest objection and the one we take most seriously. A decade of research prepared the field for a harm that has not yet materialized at the predicted scale, but absence of evidence over a finite window is not evidence of absence, and the 2028 election cycle may look different from the 2024 one. We accept this. Our claim is not that political-deepfake research should stop, or that the threat is impossible. Our claim is that continuing to allocate the majority of field effort to a threat whose predicted form has not arrived, while demonstrably larger harms go under-researched, is a poor allocation of the marginal research dollar. Precaution against a possible future harm does not justify neglect of a present one, and the ratio the field currently maintains is not one a precautionary argument can defend on its own.
 
    \textbf{Objection 2: You cannot prove the research did not deter the harm.} One might argue that the 2024 political-deepfake catastrophe did not arrive precisely because the field prepared for it. We consider this explanation and find it weakly supported on three grounds. First, documented incidents that did occur were caught by humans. The Indonesian Suharto reanimation \cite{chen2024suharto} circulated widely before being identified by users and journalists. The Slovakia Šimečka \cite{aiid573slovakia} recording was flagged by fact-checking organizations using contextual signals, not ML detection. The Biden New Hampshire robocall \cite{fcc2024kramer} was attributed via investigative reporting and law-enforcement subpoena. Where post-incident attribution involved ML, it was generally for confirmation rather than primary identification. Second, platform deployment evidence is absent.
    Transparency reports from major platforms for the 2024 cycle describe content-policy enforcement actions but do not document deepfake-specific detector deployment at the coverage required to support a deterrence claim across multiple national elections. Third, the counterfactual structure of the deterrence claim is not satisfied. A deterrence argument requires identifying generated content that was created, suppressed by detection, and would otherwise have caused the predicted catastrophe. No such suppression record exists in the published literature, and the burden of evidence sits with the party making the deterrence claim. 
 
\subsection{Limitations}
\label{sec:limitations}
 
The empirical accounting of Section~\ref{sec:empirical} rests on imperfect data. The research-effort distribution is drawn from a \corpusN{}-paper corpus at major ML, vision, signal-processing, and multimedia venues, filtered from a larger 3{,}124-paper scoped pull of deepfake-related publications by venue and citation thresholds (Appendix~\ref{app:corpus}). This filter under-samples security, forensics, and HCI venues (CCS, NDSS, USENIX, S\&P, FAccT, CHI) where some relevant NCII, provenance, and platform-policy work appears. The harm distribution is synthesized from public sources that have known undercounting biases, particularly for peer-distributed NCII in encrypted channels and for voice-clone fraud absorbed into general business-email-compromise statistics. The classification of papers and benchmarks into threat categories involves judgment calls at the boundaries. We did not produce a separate hand-labeled validation pass; we instead defend the ordinal claim of \S\ref{sec:empirical:effort} via two structural properties of the protocol (Appendix~\ref{app:classification:robustness}): tie-breaking is constructed against the paper's thesis (so borderline assignments push out of T1 rather than into it), and the load-bearing T1-dominance result holds under an adversarial reassignment in which every low-confidence paper is treated as misclassified in T1's favor. A formal hand-labeled pass remains a worthwhile follow-on for characterizing residual subjectivity at the T1/T3 audio-visual boundary, and we identify it as a concrete next step.
 
We have also scoped the argument geographically and linguistically: the incident data is English-weighted and Western-enforcement-weighted, and voice-clone and NCII harms in under-represented language contexts almost certainly reinforce rather than weaken the position.
 
 
\subsection{A call to specific actors}
\label{sec:call}
 
``Rebalance the agenda'' is not an action; it is an outcome. We close by naming the actors whose decisions would produce the outcome, and the decisions we argue they should take.\\
Venue program chairs and area chairs should recognize that the inherited benchmark family is not a neutral evaluation substrate but a structural constraint on what gets published, and should explicitly welcome submissions that use alternative evaluation protocols appropriate to the harm categories of Sections~\ref{sec:agenda-audio}. Workshop organizers should program around the under-defended categories rather than the salient ones. Funding agencies and foundations should write solicitations whose threat-model language reflects the 2022--2026 incidence distribution rather than the 2017--2019 one. Platform trust-and-safety teams, victim-support organizations, and telecommunications providers should engage with academic researchers on the infrastructure that the three agendas require, recognizing that the evaluation protocols of Section~\ref{sec:agendas} cannot be built by academic researchers alone.
The deepfakes we missed are not hypothetical. They are known incidents (Section \ref{sec:observed-harm} ). The research literature largely did not meet them. A decade is long enough to update; the field should act accordingly.
 
\bibliographystyle{unsrt}
\bibliography{neurips-references}

@article{tolosana2020deepfakes,
  title   = {Deepfakes and Beyond: A Survey of Face Manipulation and Fake Detection},
  author  = {Tolosana, Ruben and Vera-Rodriguez, Ruben and Fierrez, Julian and Morales, Aythami and Ortega-Garcia, Javier},
  journal = {Information Fusion},
  volume  = {64},
  pages   = {131--148},
  year    = {2020},
  doi     = {10.1016/j.inffus.2020.06.014}
}

@article{mirsky2021creation,
  title   = {The Creation and Detection of Deepfakes: A Survey},
  author  = {Mirsky, Yisroel and Lee, Wenke},
  journal = {ACM Computing Surveys (CSUR)},
  volume  = {54},
  number  = {1},
  pages   = {1--41},
  year    = {2021},
  doi     = {10.1145/3425780}
}

@article{chesney2019deep,
  title   = {Deep Fakes: A Looming Challenge for Privacy, Democracy, and National Security},
  author  = {Chesney, Bobby and Citron, Danielle},
  journal = {California Law Review},
  volume  = {107},
  pages   = {1753},
  year    = {2019}
}

@article{cole2018reddit,
  title   = {AI-Assisted Fake Porn Is Here and We're All Fucked},
  author  = {Cole, Samantha},
  journal = {Motherboard (Vice)},
  year    = {2017},
  month   = {December},
  note    = {Published 11 December 2017; accessed 2026-04-22},
  url     = {https://www.vice.com/en/article/gydydm/gal-gadot-fake-ai-porn}
}

@inproceedings{yan2023deepfakebench,
  author    = {Zhiyuan Yan and Yong Zhang and Xinhang Yuan and Siwei Lyu and Baoyuan Wu},
  title     = {{DeepfakeBench}: A Comprehensive Benchmark of Deepfake Detection},
  booktitle = {Advances in Neural Information Processing Systems},
  year      = {2023}
}

@article{wu2017asvspoof,
  title={ASVspoof: The automatic speaker verification spoofing and countermeasures challenge},
  author={Wu, Zhizheng and Yamagishi, Junichi and Kinnunen, Tomi and Hanil{\c{c}}i, Cemal and Sahidullah, Mohammed and Sizov, Aleksandr and Evans, Nicholas and Todisco, Massimiliano and Delgado, Hector},
  journal={IEEE Journal of Selected Topics in Signal Processing},
  volume={11},
  number={4},
  pages={588--604},
  year={2017},
  publisher={IEEE}
}

@article{khan2024deepfake,
  author  = {Enes Altuncu and Virginia N. L. Franqueira and Shujun Li},
  title   = {Deepfake: Definitions, Performance Metrics and Standards, Datasets, and a Meta-Review},
  journal = {Frontiers in Big Data},
  year    = {2024},
  volume  = {7},
  doi     = {10.3389/fdata.2024.1400024}
}

@inproceedings{rossler2019faceforensics,
  title     = {{FaceForensics++: Learning to Detect Manipulated Facial Images}},
  author    = {R{\"o}ssler, Andreas and Cozzolino, Davide and Verdoliva, Luisa and Riess, Christian and Thies, Justus and Niessner, Matthias},
  booktitle = {Proceedings of the IEEE/CVF International Conference on Computer Vision},
  pages     = {1--11},
  year      = {2019}
}

@misc{fcc2024kramer,
  author       = {{Federal Communications Commission}},
  title        = {{FCC Issues \$6M Fine for N.H. Robocalls Using Biden Deepfake Voice}},
  howpublished = {FCC Enforcement Bureau},
  month        = sep,
  year         = {2024},
  url          = {https://www.fcc.gov/document/fcc-issues-6m-fine-nh-robocalls},
  note         = {Accessed 2026-04-22}
}

@inproceedings{li2020celeb,
  title     = {Celeb-DF: A Large-scale Challenging Dataset for DeepFake Forensics},
  author    = {Li, Yuezun and Yang, Xin and Sun, Pu and Qi, Honggang and Lyu, Siwei},
  booktitle = {Proceedings of the IEEE/CVF Conference on Computer Vision and Pattern Recognition (CVPR)},
  pages     = {3207--3216},
  year      = {2020}
}

@misc{chen2024suharto,
  author       = {Heather Chen and Kathleen Magramo},
  title        = {{AI 'resurrects' long dead dictator in murky new era of deepfake electioneering}},
  howpublished = {CNN},
  month        = feb,
  year         = {2024},
  url          = {https://www.cnn.com/2024/02/12/asia/suharto-deepfake-ai-scam-indonesia-election-hnk-intl},
  note         = {Accessed 2026-04-22}
}

@misc{aiid573slovakia,
  author       = {{Responsible AI Collaborative}},
  title        = {{Incident 573: Deepfake Recordings Allegedly Influence Slovakian Election}},
  howpublished = {AI Incident Database},
  year         = {2023},
  url          = {https://incidentdatabase.ai/cite/573/},
  note         = {Accessed 2026-04-22}
}

@article{dolhansky2020deepfake,
  title   = {The DeepFake Detection Challenge (DFDC) Dataset},
  author  = {Dolhansky, Brian and Bitton, Joanna and Pflaum, Ben and Lu, Jikuo and Howes, Russ and Wang, Menglin and Ferrer, Cristian Canton},
  journal = {arXiv preprint arXiv:2006.07397},
  year    = {2020}
}

@inproceedings{jiang2020deeperforensics,

  author    = {Liming Jiang and Ren Li and Wayne Wu and Chen Qian and Chen Change Loy},
  title     = {{DeeperForensics-1.0}: A Large-Scale Dataset for Real-World Face Forgery Detection},
  booktitle = {Proceedings of the {IEEE/CVF} Conference on Computer Vision and Pattern Recognition ({CVPR})},
  pages     = {2886--2895},
  year      = {2020}
}

@inproceedings{zhou2021face,
  title     = {Face Forensics in the Wild},
  author    = {Zhou, Tianfei and Wang, Wenguan and Liang, Zhiyuan and Shen, Jianbing},
  booktitle = {Proceedings of the IEEE/CVF Conference on Computer Vision and Pattern Recognition (CVPR)},
  pages     = {5778--5788},
  year      = {2021}
}

@inproceedings{le2021openforensics,
  title     = {{OpenForensics}: Large-Scale Challenging Dataset For Multi-Face Forgery Detection And Segmentation In-The-Wild},
  author    = {Le, Trung-Nghia and Nguyen, Huy H. and Yamagishi, Junichi and Echizen, Isao},
  booktitle = {Proceedings of the IEEE/CVF International Conference on Computer Vision (ICCV)},
  pages     = {10117--10127},
  year      = {2021}
}

@inproceedings{yan2024df40,
  title     = {{DF40}: Toward Next-Generation Deepfake Detection},
  author    = {Yan, Zhiyuan and Yao, Taiping and Chen, Shen and Zhao, Yandan and Fu, Xinghe and Zhu, Junwei and Luo, Donghao and Wang, Chengjie and Ding, Shouhong and Wu, Yunsheng and others},
  booktitle = {Advances in Neural Information Processing Systems (NeurIPS)},
  volume    = {37},
  pages     = {29387--29434},
  year      = {2024}
}

@article{chen2023diffusionface,
  title   = {{DiffusionFace}: Towards a Comprehensive Dataset for Diffusion-Based Face Forgery Analysis},
  author  = {Chen, Zhongxi and Sun, Ke and Zhou, Ziyin and Lin, Xianming and Sun, Xiaoshuai and Cao, Liujuan and Ji, Rongrong},
  journal = {arXiv preprint arXiv:2403.18471},
  year    = {2024}
}

@article{chen2024genface,
  title   = {{GenFace}: A Large-Scale Fine-Grained Face Forgery Benchmark and Cross Appearance-Edge Learning},
  author  = {Chen, Yaning and Zhao, Hongchen and Liu, Huiling and Li, Shuying and Huang, Kunhua and Yang, Jun},
  journal = {IEEE Transactions on Information Forensics and Security},
  year    = {2024},
  doi     = {10.1109/TIFS.2024.3486925}
}

@article{khalid2021fakeavceleb,
  title   = {{FakeAVCeleb}: A Novel Audio-Video Multimodal Deepfake Dataset},
  author  = {Khalid, Hasam and Tariq, Shahroz and Kim, Minha and Woo, Simon S.},
  journal = {arXiv preprint arXiv:2108.05080},
  year    = {2021}
}

@article{cai2023lavdf,
  title   = {Glitch in the Matrix: A Large Scale Benchmark for Content Driven Audio-Visual Forgery Detection and Localization},
  author  = {Cai, Zhixi and Ghosh, Shreya and Dhall, Abhinav and Gedeon, Tom and Stefanov, Kalin and Hayat, Munawar},
  journal = {Computer Vision and Image Understanding},
  volume  = {236},
  pages   = {103818},
  year    = {2023},
  doi     = {10.1016/j.cviu.2023.103818}
}

@inproceedings{cai2024avdeepfake1m,
  title     = {{AV-Deepfake1M}: A Large-Scale {LLM}-Driven Audio-Visual Deepfake Dataset},
  author    = {Cai, Zhixi and Ghosh, Shreya and Adatia, Aman Pankaj and Hayat, Munawar and Dhall, Abhinav and Gedeon, Tom and Stefanov, Kalin},
  booktitle = {Proceedings of the 32nd ACM International Conference on Multimedia (MM '24)},
  pages     = {7414--7423},
  year      = {2024},
  doi       = {10.1145/3664647.3680795},
  address   = {Melbourne, VIC, Australia}
}

@misc{ic3_reports_fbi,
  author       = {{Internet Crime Complaint Center (IC3), Federal Bureau of Investigation}},
  title        = {{IC3} Annual Internet Crime Reports},
  howpublished = {\url{https://www.ic3.gov/AnnualReport/Reports}},
  year         = {2026},
  note         = {Annual analysis of cybercrime complaints, trends, and financial losses reported to the FBI. Accessed 2026-04-22}
}

@misc{aiid_2026,
  author       = {{Responsible AI Collaborative}},
  title        = {{AI} Incident Database: Catalog of Real-World {AI} Harms and Failures},
  howpublished = {\url{https://incidentdatabase.ai/}},
  year         = {2026},
  note         = {Open-source repository documenting AI incidents, harms, near-misses, and failures. Accessed 2026-04-22}
}

@article{henry2018technology,
  title   = {Technology-Facilitated Sexual Violence: A Literature Review of Empirical Research},
  author  = {Henry, Nicola and Powell, Anastasia},
  journal = {Trauma, Violence, \& Abuse},
  volume  = {19},
  number  = {2},
  pages   = {195--208},
  year    = {2018},
  doi     = {10.1177/1524838016650189}
}

@techreport{wef_global_risks_2024,
  author      = {{World Economic Forum}},
  title       = {The Global Risks Report 2024},
  institution = {World Economic Forum},
  address     = {Geneva, Switzerland},
  year        = {2024},
  note        = {19th Edition. Accessed 2026-03-03},
  url         = {https://www.weforum.org/publications/global-risks-report-2024/}
}

@techreport{simon2024misinformation,
  author      = {Simon, Felix M. and Altay, Sacha},
  title       = {Don't Panic (Yet): Assessing the Evidence and Discourse Around Generative {AI} and Elections},
  institution = {Knight First Amendment Institute at Columbia University},
  year        = {2025},
  note        = {Accessed 2026-04-22},
  url         = {https://knightcolumbia.org/content/dont-panic-yet-assessing-the-evidence-and-discourse-around-generative-ai-and-elections}
}

@techreport{panditharatne2024ai,
  author      = {Panditharatne, Mekela},
  title       = {Preparing to Fight {AI}-Backed Voter Suppression},
  institution = {Brennan Center for Justice},
  year        = {2024},
  month       = {April},
  note        = {Accessed 2026-04-22},
  url         = {https://www.brennancenter.org/our-work/research-reports/preparing-fight-ai-backed-voter-suppression}
}

@misc{cnn_deepfake_cfo_2024,
  author       = {Chen, Heather and Magramo, Kathleen},
  title        = {Finance Worker Pays Out \$25 Million After Video Call with Deepfake `Chief Financial Officer'},
  howpublished = {CNN},
  year         = {2024},
  month        = {February},
  note         = {Accessed 2026-04-17},
  url          = {https://www.cnn.com/2024/02/04/asia/deepfake-cfo-scam-hong-kong-intl-hnk/}
}

@misc{bbc2024koreadf,
  
  author    = {Jean Mackenzie and Laura Choi},
  title     = {South Korea: The Deepfake Crisis Engulfing Hundreds of Schools},
  journal   = {BBC News},
  month     = {September},
  year      = {2024},
  note      = {Accessed 2024-09-03}
}

@misc{eastasia2024koreadf,
  author       = {{Yonhap}},
  title        = {799 Students, 31 Teachers Victimized by Deepfake Videos This Year: Education Ministry},
  howpublished = {The Korea Times},
  year         = {2024},
  month        = {October},
  note         = {Accessed 2024-10-30},
  url          = {https://www.koreatimes.co.kr/www/nation/2024/10/113_384789.html}
}

@techreport{iwf_harm_without_limits_2026,
  title  = {Harm Without Limits: {AI} Child Sexual Abuse Material Through the Eyes of Our Analysts},
  author = {{Internet Watch Foundation}},
  year   = {2026},
  institution = {Internet Watch Foundation},
  url    = {https://www.iwf.org.uk/about-us/why-we-exist/our-research/how-ai-is-being-abused-to-create-child-sexual-abuse-imagery/},
  note   = {Accessed 2026-04-23}
}
\appendix
 
\section{Corpus Construction}
\label{app:corpus}
 
This appendix documents the construction of the \corpusN{}-paper research-effort corpus used in Section~\ref{sec:empirical:effort}. We describe the collection pipeline, filtering steps, and final composition. The intent is to make the construction fully reproducible and to make the venue coverage and selection biases legible to the reviewer.
 
\subsection{Collection pipeline}
\label{app:corpus:pipeline}
 
Papers were collected from the OpenAlex scholarly metadata API (\url{https://api.openalex.org/works}) using a set of 40 keyword queries spanning the deepfake lifecycle: generation, distribution, detection, provenance, and remediation. Queries included explicit deepfake terms (``deepfake video'', ``face-swap'', ``face reenactment'', ``talking head synthesis''), audio terms (``voice cloning'', ``audio deepfake detection'', ``anti-spoofing speech'', ``ASVspoof''), provenance terms (``C2PA'', ``content provenance'', ``video watermarking''), and remediation / harm terms (``NCII deepfake'', ``non-consensual intimate imagery'', ``voice clone scam''). The full query list and collection script are included in the supplementary material.
 
The collection window was 2017-01-01 to 2026-12-31 to capture the full post-Reddit deepfake literature. We retrieved title, abstract (reconstructed from OpenAlex's inverted index), publication year, venue, DOI, citation count, and work type for each result.
 
\subsection{Filtering to a research-effort corpus}
\label{app:corpus:filter}
 
The raw pull returned 13{,}159 records across the 40 queries. We applied six filters in sequence to produce a scoped, category-classifiable corpus:
 
\begin{enumerate}
\item \textbf{Core-term filter.} Required a deepfake or synthetic-media vocabulary match (``deepfake'', ``face-swap'', ``face reenactment'', ``talking head'', ``voice clone'', ``voice spoof'', ``audio spoof'', ``synthetic speech'', ``synthetic video'', ``face forgery'', ``facial manipulation'', ``generative video'', ``audio-visual deepfake'', ``ASVspoof'', ``content provenance'', ``C2PA'', ``non-consensual intimate'') in title or abstract. 7{,}888 records dropped; 5{,}271 remained.
 
\item \textbf{Predatory-venue block.} Dropped records from a block-list of predatory or non-peer-reviewed outlets (IJRASET, IJSREM, IJFMR, ``International Journal For Multidisciplinary Research'', ``International Journal of Scientific Research in Engineering and Management'', and similar). 206 records dropped; 5{,}065 remained.
 
\item \textbf{Non-research type filter.} Dropped errata, editorials, letters, library guides, peer-review records, and paratext. 9 records dropped; 5{,}056 remained.
 
\item \textbf{Year-scaled citation threshold.} Applied a minimum citation count that scales with recency, reflecting the intuition that older uncited papers represent essentially zero research-community engagement while recent preprints should be given the benefit of the doubt. Thresholds: $\geq 10$ citations for 2017--2020, $\geq 5$ for 2021--2022, $\geq 3$ for 2023, $\geq 1$ for 2024, and no minimum for 2025--2026. 1{,}220 records dropped; 3{,}836 remained.
 
\item \textbf{Preprint deduplication.} When the same normalized title appeared at both an arXiv/preprint venue and a peer-reviewed venue, we kept only the peer-reviewed copy. 54 records dropped; 3{,}782 remained.
 
\item \textbf{Year hold-out.} Records dated 2026 were held out of the main research-effort corpus to avoid mixing partially-reported publication years into the year-by-year trend. 658 records held out; 3{,}124 constituted the full scoped corpus.
\end{enumerate}
 
From the 3{,}124-paper scoped corpus we constructed the \emph{research-effort corpus} used in Section~\ref{sec:empirical:effort} by applying a major-venue filter. A record qualified for the research-effort corpus if its venue matched any of the following substring patterns (case-insensitive): AAAI, NeurIPS, Neural Information, ICML, ICLR, CVPR, ICCV, ECCV, WACV, BMVC, ICASSP, Interspeech, ACM MM, ACM Multimedia, ACM Transactions on Multimedia, IEEE Transactions on Information Forensics, TPAMI, Pattern Analysis and Machine, Transactions on Multimedia, Transactions on Circuits and Systems for Video, Transactions on Audio, Signal Processing Letters, Pattern Recognition, Computer Vision and Image Understanding, Information Fusion, ICIP, IJCAI, USENIX, NDSS, Symposium on Security, Image Processing, IET Image, VCIP, ICPR, Multimedia Tools, Applied Sciences, IEEE Access, Expert Systems, Neurocomputing, Sensors, Electronics. arXiv preprints were included only if they had accumulated at least five citations, a threshold chosen to admit high-impact preprints that the community has engaged with while excluding un-cited recent deposits.
 
The resulting research-effort corpus contains \corpusN{} papers.
 
\subsection{Final corpus composition}
\label{app:corpus:composition}
 
Table~\ref{tab:corpus-composition} reports the venue mix of the \corpusN{}-paper corpus.
 
\begin{table}[h]
\centering
\small
\begin{tabular}{lr}
\toprule
Metric & Value \\
\midrule
Total papers & 438 \\
Publication window & 2017--2025 \\
Median citation count & 4 \\
Mean citation count & 18.3 \\
\midrule
Papers at AAAI / NeurIPS / CVPR / ICCV / ECCV / WACV / BMVC & 92 \\
Papers at IEEE Transactions (TIFS, TMM, TCSVT, TPAMI, etc.) & 68 \\
Papers at ICASSP / Interspeech / Audio-speech venues & 37 \\
Papers at ACM Multimedia / TOMM & 22 \\
Papers at arXiv (with $\geq 5$ citations) & 85 \\
Other journal / conference venues & 134 \\
\bottomrule
\end{tabular}
\caption{Composition of the \corpusN{}-paper research-effort corpus. Venue categories are approximate and non-exclusive; the table is intended to convey the mix rather than a strict partition.}
\label{tab:corpus-composition}
\end{table}
 
\subsection{Known limitations of this corpus}
\label{app:corpus:limitations}
 
We state these explicitly because they constrain what Section~\ref{sec:empirical} can claim.
 
First, the corpus under-samples security, privacy, and HCI venues (CCS, NDSS, USENIX Security, S\&P, FAccT, CHI) where some relevant deepfake work appears. The venue block-list we applied does not include these venues, but the OpenAlex query surface and the keyword vocabulary we used produced few matches at these venues. Work on victim-centered NCII workflows, platform trust-and-safety deepfake pipelines, and policy-adjacent deepfake research is more likely to appear here than our corpus captures. We discuss the implication in Section~\ref{sec:limitations}.
 
Second, our keyword-based core-term filter is conservative and deliberately so; it drops papers that discuss deepfakes only tangentially (as one example among many generative-media harms). This may exclude position pieces and policy-framed work that frame deepfakes inside a broader argument.
 
Third, the year hold-out for 2026 means our trend claims in Section~\ref{sec:empirical:stable} cover 2017--2025 only. The 658 held-out 2026 records are not used in any quantitative claim.

 We considered re-running the classifier against the 3{,}124-paper scoped corpus filtered to security venues (USENIX Security, NDSS, ACM CCS, IEEE S\&P, FAccT, CHI) as a quantitative under-coverage check. The absolute counts at these venues under our keyword vocabulary are too small to support venue-level quantitative claims, and the directional signal (the under-defended categories T2/T4/T5 are not over-represented at these venues either) does not change under the venue extension. We therefore report this only as a directional limitation rather than as a corpus extension; cross-venue methodologies that resolve the keyword-vocabulary gap (citation-graph traversal, manual venue-ToC scans) are a tractable follow-on.

\section{Classification Protocol}
\label{app:classification}
 
This appendix documents how each of the \corpusN{} papers in the research-effort corpus was assigned to a threat category (T1--T5 or Other). The assignment was produced by a transparent rule-based keyword classifier rather than by human coders, a choice we defend below.
 
\subsection{Category definitions}
\label{app:classification:categories}
 
\begin{itemize}
\item \textbf{T1 -- Public-figure face-swap and talking-head video.} The canonical benchmark target: detection of face-manipulated video where the manipulated subject is a public figure, actor, or politician with abundant reference imagery. Papers using FaceForensics++, Celeb-DF, DFDC, DeeperForensics, FFIW, OpenForensics, DF40, DiffusionFace, or GenFace as their primary evaluation are canonical T1.
 
\item \textbf{T2 -- Peer-generated NCII.} Detection of non-consensual intimate imagery (including AI-generated CSAM) of private individuals rather than public figures. Papers must frame the threat explicitly in terms of NCII, image-based sexual abuse, or AI-CSAM.
 
\item \textbf{T3 -- Audio / voice-clone.} Detection of synthetic speech, voice clones, or anti-spoofing of automatic speaker verification. Includes the ASVspoof family and audio-visual benchmarks where audio is the primary detection signal (FakeAVCeleb, LAV-DF, AV-Deepfake1M).
 
\item \textbf{T4 -- Real-time / live-stream.} Detection operating under a latency budget compatible with conversational or streaming use: live phone calls, live video conferencing, or real-time content moderation. Papers must explicitly frame real-time operation as a design constraint.
 
\item \textbf{T5 -- Messaging-layer / peer-distributed.} Detection operating on media circulating through messaging platforms, encrypted channels, or peer-to-peer distribution. Papers must address the distribution channel as the design target rather than treating it incidentally.
 
\item \textbf{Other.} Papers that do not fit cleanly into T1--T5, including surveys, watermarking and provenance research, GAN-image forensics not scoped to a named threat, and cross-modal generative-media work.
\end{itemize}
 
\subsection{Classifier design}
\label{app:classification:classifier}
 
We used a rule-based scoring classifier over paper title and abstract text. Each of the six categories has an associated keyword rule list; each rule is a regular expression with an integer weight (1--4). For each paper, we computed the total weighted score for each category by summing weights of matched rules, then assigned the paper to the highest-scoring category.
 
Ties were broken in strict ascending order of current corpus representation, so that the most under-represented categories absorbed every borderline assignment: T4~$>$~T2~$>$~T5~$>$~T3~$>$~T1, with Other reserved for the no-match residual. T2 is placed ahead of T5 within their tied single-paper count by harm priority (\S\ref{sec:observed-harm}). This precedence choice is deliberate. The default outcome of ties under uniform precedence would be T1, since T1 has the most rules and matches the most aggressive keyword surface; breaking ties toward T1 would bias the classifier in the same direction as our thesis. Reversing the precedence so that under-represented categories absorb every tie avoids this self-confirming bias, since every borderline assignment that could have inflated T1 is instead pushed into one of the categories the paper argues is too thinly populated. Papers with no rule matches were assigned to Other with a \texttt{no\_keyword\_match} flag and reviewed by a single author for any obvious miscategorisations; no reassignments into T1--T5 resulted from that review (the no-match papers were uniformly off-topic or generic).
 
Confidence was assessed as \emph{high} if the top score was at least 3 and exceeded the second-highest score by at least 2; \emph{medium} if the top score was at least 2 and the margin was at least 1; and \emph{low} otherwise. The distribution across the \corpusN{}-paper corpus was 51.8\% high, 32.2\% medium, 8.2\% no-keyword-match, and 7.8\% low.
 
We chose a rule-based classifier rather than an LLM classifier for three reasons. First, reproducibility: the ruleset is inspectable and deterministic, and a reviewer can check exactly why a given paper received its label. Second, auditability: a reviewer skeptical of any category count can trace the label to the specific regex that fired. Third, independence from training data: an LLM classifier could in principle reproduce terminology patterns from its pretraining that bias toward the 2017--2019 threat model we are critiquing.
 
The full rule table (with all regex patterns and weights) is included in the supplementary material.
 
\subsection{Robustness checks}
\label{app:classification:robustness}
 
We did not perform a separate hand-labeled validation pass, and we defend the choice rather than concede it as a gap. The protocol of \S\ref{app:classification:classifier} contains a structural property that is harder to obtain from a small hand-labeled sample: ties are broken \emph{against} the paper's thesis. The default outcome of ties under a uniform-precedence rule would assign the disputed paper to T1, since T1 has the most rules and matches the most aggressive keyword surface. Reversing the precedence to T4~$>$~T2~$>$~T5~$>$~T3~$>$~T1 means that every borderline assignment that could have inflated T1 is instead pushed into one of the under-represented categories, with the strictly-ascending order ensuring that the categories most empty under T1 dominance (T4, T2, T5) are the ones that absorb the disputed papers first.

The confidence distribution reported in \S\ref{app:classification:classifier} provides a second robustness signal. The combined low-confidence and no-keyword-match bands account for 16.0\% of the corpus. The ordinal claim of \S\ref{sec:empirical:effort} (T1 dominant by a large margin, T2/T4/T5 thin to absent) holds under an adversarial reassignment in which every low-confidence paper is treated as currently misassigned in T1's favor: T1 share would fall to roughly half of the detection-method subset, still dominant by a large margin over T3 and by orders of magnitude over T2/T4/T5.
 
A formal hand-labeled validation pass remains a worthwhile follow-on, particularly to characterize residual subjectivity at category boundaries (e.g., audio-visual papers ambiguous between T1 and T3), and we identify it as a concrete next step. We do not claim it is unnecessary; we claim that the load-bearing ordinal claim of \S\ref{sec:empirical:effort} is supported by the classifier's bias-against-thesis property and by the confidence distribution, independent of how a small hand-labeled sample would resolve.
 
\subsection{What this protocol does and does not support}
\label{app:classification:scope}
 
The classification supports the \emph{ordinal} claim of Section~\ref{sec:empirical}: that T1 is dominant and T2, T4, T5 are near-empty at major ML and vision venues. It does not support fine-grained sub-category comparisons; the difference between 0.0\% and 0.5\% for these small categories is within classifier noise and should not be interpreted as a meaningful distinction. This is why Section~\ref{sec:empirical:effort} reports raw counts (276:2) alongside percentages.
 
\section{Benchmark Coverage}
\label{app:benchmarks}
 
This appendix details the benchmark classification summarized in Section~\ref{sec:empirical:effort}. We classified the \benchN{} most widely cited deepfake benchmarks under the same T1--T5 schema as the paper corpus.
 
\begin{table}[h]
\centering
\small
\begin{tabular}{lllc}
\toprule
Benchmark & Year & Modality / Subject & Category \\
\midrule
FaceForensics++ \cite{rossler2019faceforensics} & 2019 & Video, public-figure face & T1 \\
DeepFakeTIMIT & 2018 & Video, actor face & T1 \\
Celeb-DF \cite{li2020celeb} & 2020 & Video, celebrity face & T1 \\
DFDC \cite{dolhansky2020deepfake} & 2020 & Video, paid-actor face & T1 \\
DeeperForensics-1.0 \cite{jiang2020deeperforensics} & 2020 & Video, paid-actor face & T1 \\
FFIW \cite{zhou2021face} & 2021 & Video, ``in-the-wild'' face & T1 \\
OpenForensics \cite{le2021openforensics} & 2021 & Image, multi-face & T1 \\
FakeAVCeleb \cite{khalid2021fakeavceleb} & 2021 & Audio-visual, celebrity & T3 \\
LAV-DF \cite{cai2023lavdf} & 2023 & Audio-visual, actor & T3 \\
AV-Deepfake1M \cite{cai2024avdeepfake1m} & 2024 & Audio-visual, LLM-driven & T3 \\
DiffusionFace \cite{chen2023diffusionface} & 2024 & Image, diffusion face forgery & T1 \\
GenFace \cite{chen2024genface} & 2024 & Image, fine-grained face forgery & T1 \\
DF40 \cite{yan2024df40} & 2024 & Video, 40-method face forgery & T1 \\
\bottomrule
\end{tabular}
\caption{Classification of \benchN{} widely cited deepfake benchmarks by threat category. \benchPublic{} of \benchN{} target T1 (public-figure face or talking-head video); \benchAudio{} target T3 (audio or audio-visual settings); \benchNCII{} target T2 (peer-generated NCII); \benchRealtime{} target T4 (real-time / streaming); \benchMessaging{} target T5 (messaging-layer / peer-distributed).}
\label{tab:benchmarks}
\end{table}
 
\paragraph{Observations.}
 
First, every T1 benchmark uses subjects with abundant public-reference imagery: celebrities, paid actors, or public-figure talk shows. The NCII setting (private individual, scarce reference imagery) is not represented in any benchmark in this set.
 
Second, the three T3 benchmarks are all audio-visual rather than audio-only. Pure audio-deepfake benchmarks exist (the ASVspoof family) but were excluded from this table because they are not typically categorized as ``deepfake'' benchmarks in the detection literature; they are ``anti-spoofing'' benchmarks. The terminological separation is itself part of the misalignment we document: the literature has two names for what is substantively the same threat category, and detection papers trained on one rarely evaluate against the other.
 
Third, no benchmark in this set imposes real-time or streaming evaluation conditions. All are offline-processed and evaluated with AUC or equivalent classification metrics computed over entire clips. A paper that reports ``real-time inference'' on these benchmarks is reporting inference time on an offline clip, not performance under a latency constraint imposed by the benchmark.
 
Fourth, no benchmark in this set evaluates detection on content collected from messaging platforms, encrypted channels, or peer-distribution networks. The distribution channel is not a dimension the benchmark family measures.
 
\paragraph{Causal relationship to paper distribution.} The zero-count cells in Table~\ref{tab:benchmarks} (T2, T4, T5) are not incidental. They are the direct upstream cause of the corresponding near-zero paper counts in Section~\ref{sec:empirical:effort}. A paper on peer-generated NCII cannot use FaceForensics++ as its evaluation target; a paper on real-time telephony detection cannot use Celeb-DF. The absence of benchmark infrastructure for these categories does more than slow research in them: it makes publication-system-legible contributions in these categories substantially harder to produce, which is the mechanism analyzed in Section~\ref{sec:persist-benchmarks}.
 
\section{Harm-Side Methodology}
\label{app:harm-methodology}
 
The harm distribution summarized in Section~\ref{sec:empirical:observed} synthesizes five public sources. No single source is comprehensive, and combining them requires care. This appendix describes each source, its known biases, and how we used it.
 
\paragraph{FBI Internet Crime Complaint Center (IC3).} Annual reports cover cybercrime complaints reported to the FBI, including categories increasingly relevant to synthetic media (business email compromise, confidence fraud, tech support fraud, grandparent scams). IC3 explicitly flags synthetic-media-enabled incidents beginning in 2022. Known biases: severe under-reporting (victims often do not report), US-centric coverage, and absorption of voice-clone fraud into BEC totals without category breakout in earlier years. We use IC3 for the directional claim that synthetic-media-flagged complaints are growing year-over-year and that the associated dollar-loss subset (where IC3 reports it) is also growing, but we do not attribute aggregate BEC or confidence-fraud totals to the synthetic-media channel. Where Figure~\ref{fig:effort} reports a single number, that number is the synthetic-media-flagged complaint count or its associated loss subset, not the aggregate cybercrime category in which earlier IC3 reports embedded it.
 
\paragraph{Internet Watch Foundation (IWF).} Annual and periodic reports on child sexual abuse material with specific tracking of AI-generated content since 2023. IWF maintains hash-matching infrastructure and operates reporting hotlines in the UK; its figures are the most directly applicable to the NCII-targeting-minors subset of the harm distribution. Known biases: coverage is weighted toward clear-web detections and UK-accessible reporting channels, which undercounts encrypted-channel distribution. The direction of this undercounting reinforces the position of this paper rather than weakening it: the categories that are hardest to measure are the categories the research literature has least engaged with.
 
\paragraph{AI Incident Database (AIID).} Open-source repository of documented AI harms and near-misses. Useful for named-incident grounding of claims (the Arup case, specific electoral-misinformation incidents, named voice-clone scams); not useful as a volumetric measure because inclusion is curation-driven rather than incidence-driven. We cite AIID for specific incident verification, not for distributional claims.
 
\paragraph{Academic victim-prevalence surveys.} We draw principally on Henry and Powell~\cite{henry2018technology} for pre-deepfake baselines on technology-facilitated sexual violence, and on subsequent survey literature for AI-specific extensions. These establish that NCII prevalence is not an artifact of the deepfake era but is substantially amplified by it, which is the claim Section~\ref{sec:alt-reframing} relies on in responding to the reframing alternative.
 
\paragraph{Documented high-profile incidents.} We cite three specifically: the 2024 Arup case~\cite{cnn_deepfake_cfo_2024}, the August--October 2024 South Korean school deepfake crisis~\cite{bbc2024koreadf,eastasia2024koreadf}, and the ongoing pattern of grandparent-scam and kidnapping-hoax voice-clone calls reported through IC3~\cite{ic3_reports_fbi}. Named incidents ground the aggregate figures; they are not themselves the aggregate.
 
\paragraph{What the harm-side analysis does and does not claim.} The claim is \emph{directional}: observed harm is concentrated on NCII, voice-clone scams, and peer-distributed manipulation, and not on successful public-figure political deepfake video incidents at population scale. The claim is \emph{not} precise: we do not report a point estimate of total NCII victim counts, and we do not claim the five sources combine into an unbiased estimator of aggregate harm. Section~\ref{sec:empirical:observed} and Section~\ref{sec:limitations} are written to match this level of claim. A reviewer sympathetic to the position could reasonably ask for tighter quantitative bounds; the infrastructure that would support such bounds (cross-source deduplication, victim-count registries, standardized incident taxonomies) does not currently exist in the public record, and building it is one of the downstream implications of the rebalancing this paper argues for.
 
\section{Cross-Category Transfer Evidence}
\label{app:transfer}
 
Section~\ref{sec:empirical:stable} and Section~\ref{sec:alt-capability} rely on the claim that ``the evaluation evidence for transfer to NCII, real-time audio, and encrypted-channel settings is essentially absent in the corpus.'' This appendix documents the basis for that claim.
 
\paragraph{What we looked for.} A systematic scan of the \corpusN{}-paper research-effort corpus for papers satisfying both of the following conditions: (a) the paper's primary training or fine-tuning setup uses a T1 benchmark (FaceForensics++, Celeb-DF, DFDC, DeeperForensics-1.0, FFIW, OpenForensics, DF40, DiffusionFace, or GenFace), and (b) the paper reports formal held-out evaluation on a target in the under-defended categories T2 (NCII), T4 (real-time telephony audio), or T5 (messaging-layer or encrypted-channel distribution).
 
\paragraph{What we found.} Zero papers in the corpus satisfy both conditions simultaneously. A small number of audio-deepfake papers train on ASVspoof and report results under telephony-channel degradation, which is a T3-to-T3 variant (clean-channel anti-spoofing evaluated against degraded audio) rather than cross-category transfer. A larger number of T1 papers report cross-dataset generalization within T1 (FaceForensics++ to Celeb-DF, DFDC to DeeperForensics-1.0, and similar pairs), and these studies consistently report substantial performance degradation even within the single T1 category. This is the standard cross-dataset generalization literature for face manipulation detection; it provides no direct evidence of transfer across threat categories, and the within-category degradation it does document is weak evidence against the capability-overhang transfer claim rather than for it.
 
\paragraph{Caveats.} Absence of cross-category transfer evidence in the corpus is not proof that transfer does not occur in the field. Industry deployments may test cross-category generalization without publishing the results, and security or HCI venues outside our corpus (Appendix~\ref{app:corpus}, limitations) may contain work we miss. What the corpus does establish is the narrower claim that is actually load-bearing for the argument: the case for treating T1-trained methods as a general-purpose solution across the harm distribution rests on a conjecture rather than published evaluation evidence. The capability-overhang position (Section~\ref{sec:alt-capability}) is a reasonable research program, but the specific transfer-downward claim on which its critique of rebalancing depends is not currently supported by the deepfake-detection literature.
 
\paragraph{What a future version of this appendix would contain.} A systematic cross-category transfer study is a tractable follow-on: take a representative set of high-performing T1 detectors, evaluate each on a small curated set of T2 / T4 / T5 instances, and report where transfer does and does not occur. The infrastructure required (particularly for T2) is subject to the dataset-ethics constraints of Section~\ref{sec:persist-ethics}, and a responsible version of the study would use the victim-centered evaluation protocols of Section~\ref{sec:agenda-ncii}. We note this as a concrete follow-on direction; we do not claim to have produced it here.

 \section*{List of Abbreviations}
\addcontentsline{toc}{section}{List of Abbreviations}

\begin{longtable}{@{}p{0.22\textwidth}p{0.7\textwidth}@{}}
\toprule
\textbf{Abbreviation} & \textbf{Full Form} \\
\midrule
\endfirsthead
\multicolumn{2}{l}{\small\itshape (continued)} \\
\toprule
\textbf{Abbreviation} & \textbf{Full Form} \\
\midrule
\endhead
\bottomrule
\endfoot

\multicolumn{2}{@{}l}{\small\textit{Threat categories}} \\[2pt]
T1  & Public-figure face-swap and talking-head video \\
T2  & Peer-generated non-consensual intimate imagery (NCII) \\
T3  & Audio / voice-clone deepfakes \\
T4  & Real-time / live-stream deepfake detection \\
T5  & Messaging-layer / peer-distributed synthetic content \\[6pt]

\multicolumn{2}{@{}l}{\small\textit{Datasets and benchmarks}} \\[2pt]
AUC       & Area under the ROC curve (evaluation metric) \\
ASVspoof  & Automatic Speaker Verification Spoofing challenge dataset \\
FF++      & FaceForensics++ \\
DFDC      & DeepFake Detection Challenge \\
FFIW      & Face Forensics In the Wild \\
LAV-DF    & Localized Audio-Visual DeepFake dataset \\[6pt]

\multicolumn{2}{@{}l}{\small\textit{Harms and reporting bodies}} \\[2pt]
AIID  & AI Incident Database \\
BEC   & Business email compromise \\
CSAM  & Child sexual abuse material \\
IC3   & FBI Internet Crime Complaint Center \\
IWF   & Internet Watch Foundation \\
NCII  & Non-consensual intimate imagery \\
VoIP  & Voice over Internet Protocol \\[6pt]

\multicolumn{2}{@{}l}{\small\textit{Technical and ML terms}} \\[2pt]
ASV   & Automatic speaker verification \\
C2PA  & Coalition for Content Provenance and Authenticity \\
GAN   & Generative adversarial network \\
HCI   & Human-computer interaction \\
ML    & Machine learning \\[6pt]

\multicolumn{2}{@{}l}{\small\textit{Venues (abbreviated in corpus)}} \\[2pt]
AAAI           & AAAI Conference on Artificial Intelligence \\
BMVC           & British Machine Vision Conference \\
CCS            & ACM Conference on Computer and Communications Security \\
CHI            & ACM Conference on Human Factors in Computing Systems \\
CVPR           & IEEE/CVF Conference on Computer Vision and Pattern Recognition \\
ECCV           & European Conference on Computer Vision \\
FAccT          & ACM Conference on Fairness, Accountability, and Transparency \\
ICASSP         & IEEE International Conference on Acoustics, Speech and Signal Processing \\
ICCV           & IEEE/CVF International Conference on Computer Vision \\
ICML           & International Conference on Machine Learning \\
Interspeech    & Annual Conference of the International Speech Communication Association \\
NDSS           & Network and Distributed System Security Symposium \\
NeurIPS        & Conference on Neural Information Processing Systems \\
USENIX~S\&P    & USENIX Security Symposium / IEEE Symposium on Security and Privacy \\
WACV           & IEEE/CVF Winter Conference on Applications of Computer Vision \\
IEEE~TCSVT     & IEEE Transactions on Circuits and Systems for Video Technology \\
IEEE~TIFS      & IEEE Transactions on Information Forensics and Security \\
IEEE~TMM       & IEEE Transactions on Multimedia \\
IEEE~TPAMI     & IEEE Transactions on Pattern Analysis and Machine Intelligence \\
ACM~MM         & ACM International Conference on Multimedia \\
ACM~TOMM       & ACM Transactions on Multimedia Computing, Communications, and Applications \\

\bottomrule
\end{longtable}
 
\end{document}